\documentclass[twocolumn,prl,superscriptaddress,floatfix,showpacs]{revtex4}
\usepackage{graphics}
\usepackage{graphicx}
\usepackage{amssymb}
\usepackage{amsmath}
\usepackage{bm}

\begin{document}
\bibliographystyle{apsrev}

\title{Non-classical nucleation in supercooled nickel}
\author{F. J. Cherne}
\affiliation{Los Alamos National Laboratory\\ DX-2, Materials Dynamics\\ Los Alamos, NM 87545}
\author{M. I. Baskes}
\affiliation{Los Alamos National Laboratory\\ MST-8 Structure and Property Relations\\ Los Alamos, NM 87545}
\author{R. B. Schwarz}
\affiliation{Los Alamos National Laboratory\\ MST-8 Structure and Property Relations\\ Los Alamos, NM 87545}
\author{S. G. Srinivasan}
\affiliation{Los Alamos National Laboratory\\ MST-8 Structure and Property Relations\\ Los Alamos, NM 87545}
\author{W. Klein}
\affiliation{Los Alamos National Laboratory\\ X-7 Material Science\\
  Los Alamos NM 87545}
\affiliation{Boston University, Department of Physics, Boston, MA 02215}

\date{\today}

\begin{abstract}
  The dynamics of homogeneous nucleation and growth of crystalline 
  nickel from the super-cooled melt is examined during rapid quenching using molecular
  dynamics and a modified embedded atom method potential. 
  The character of the critical nuclei of the crystallization 
  transition is examined using common neighbor analysis and visualization. 
  At nucleation the saddle point droplet consists of randomly
  stacked planar structures with an in plane triangular order. These results
are consistent with previous theoretical results that predict that
  the nucleation process in some metals is non-classical due to the 
  presence of long-range forces and a spinodal. 

\end{abstract}

\pacs{ }

\maketitle

\section{Introduction}
Solidification of liquids usually begins by the formation of small
clusters or nuclei. 
Understanding the structure of nuclei and the dynamics of nucleation
and growth is of great scientific and technical interest. Nucleation
is a highly non-linear process and the interplay between the
non-linearity and a change in symmetry is not understood in, for
example,  the
crystallization of super-cooled liquids. In addition to the 
interesting scientific questions such as the structure of the
critical or nucleating droplet and the nature of the initial growth,  
the form of the nucleation process can have a profound effect on the
morphology of the final state. Classical nucleation
theories\cite{gunt} predict that the interior of the critical droplet
will have the same structure as the stable phase. However    
Klein and Leyvraz(KL)\cite{kle86,kle01} have argued    
that the nucleating droplet 
in supercooled liquids with long range potentials 
is non-classical and can have a
structure which is significantly different than
that of the stable phase due to the presence of a spinodal. 
They suggest that metals with effective long range interactions 
due to elastic forces would be a candidate for exhibiting this 
process which takes place arbitrarily close to a spinodal. 
The non-classical droplet is predicted to have a density close to that of  
the supercooled liquid and a triangular structure in 2 dimensions. 
In 3 dimensions either randomly stacked triangular planes or a bcc
structure is predicted depending on the details of the potential. 
In the initial growth phase of the droplet it is predicted that a core
will develop that has the density, and symmetry, of the stable solid
phase, or of a metastable solid phase.\cite{kle01,uk85}

Simple models of elastic systems exhibiting an Austenite to Martensite
transition have indicated a non-classical nucleation pattern.\cite{kle02} 
However, there has been no test of the prediction of non-classical 
nucleation in either
experiment or simulations with realistic metal potentials.
Here we present a computational study examining the nucleation and
 growth in 
metastable liquid nickel using state-of-the-art semi-empirical potentials. 
Our results indicate that the nucleation is non-classical and is
consistent with nucleation near a spinodal. It is important to stress
that by nucleating or critical droplet we refer to the saddle point
structure. Physically this is the droplet that just reaches the
critical size. This appears in the theory as a saddle
point.\cite{kle01, ch, lang}

\section{Method}
To perform our simulations we use the modified embedded atom method 
(MEAM) potential, referred to as 
potential 4 in Ref. \cite{bas97}. We extended the radial cutoff, 
r$_{cut}$, from 4.0 to 4.8 \AA to conserve energy in the liquid phase.  This
alteration does not affect the solid properties.  A thorough discussion 
of the MEAM formalism can be found in the literature \cite{bas97,bas92,bas94} and
will not be repeated here.  Recently, this particular potential was
shown to describe the properties of liquid nickel quite well
\cite{che02}. The calculated melting point was within 9 \% of
experiment.  The calculated viscosity of 3.84 mPa-s at 1775 K was in
agreement with the lower of the experimental range of values
$\sim$4.0-6 mPa-s. 

We perform a molecular dynamics quenching sequence using an
isobaric-isothermal (NPT-ensemble). The MD calculations use a time
step of 1 fs. Constant temperature is maintained by using a
Nos\'{e}-Hoover \cite{nos91} thermostat.  The Parinello-Rahman method
for pressure control is used maintaining the pressure to within
$\pm$100 bar.  The sequence is begun by equilibrating a 3D periodic
cell of (e.g.) 1500 atoms in the liquid state at a temperature slightly above
the melting point of the system for 25 ps.  The samples are cooled
at 10 K increments and held at that temperature for the length of time
corresponding to the desired cooling rate.  For example, a 10$^{12}$
K/s cooling rate would be obtained by holding the sample at each
temperature step for 10 ps.  The quench rates for this study ranged
from 10$^{11}$ K/s to 10$^{13}$ K/s.  A few calculations are
performed using a larger cell of 2048 and 12000 atoms.  

Averages of the potential energy and volume are determined at each
temperature after 1 ps.  The only exception to this is for the quench
rate of 10$^{13}$ K/s, where the average temperature and volume are
determined for the last 0.5 ps.  From this data, we determine whether
the specimen is undergoing a liquid-to-crystal or a liquid-to-amorphous
transition.  The crystallization produces an inflection on the cooling
curve and the inflection point determines the solidification
temperature, T$_{m}^{'}$.  The liquid-to-amorphous transition produces
a change of slope in the cooling curve.  The glass transition
temperature, T$_g$, is determined from the intersection of straight
lines fit to the cooling curve above and below T$_g$.

We use common neighbor analysis (CNA) to identify the local
crystalline environment of each atom \cite{fak94,bar84,hon87,jon88}.
This method provides
an efficient way to identify the local environment of each atom
in the system. The results are quite insensitive to small
displacements of the atoms. Since our CNA is not configured
to identify planar structures
we also perform a visual analysis of 
possible planar structures both surrounding the 3D Bravais lattice 
identified by the CNA and free standing(i.e. without the Bravais core).

\section{Results and Discussion}

\begin{figure}
\includegraphics{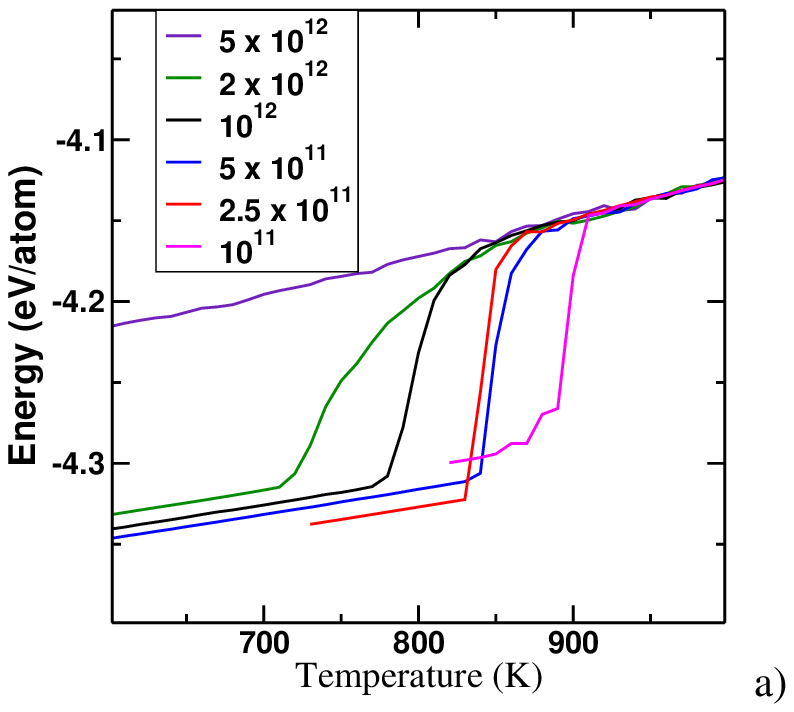}
\includegraphics{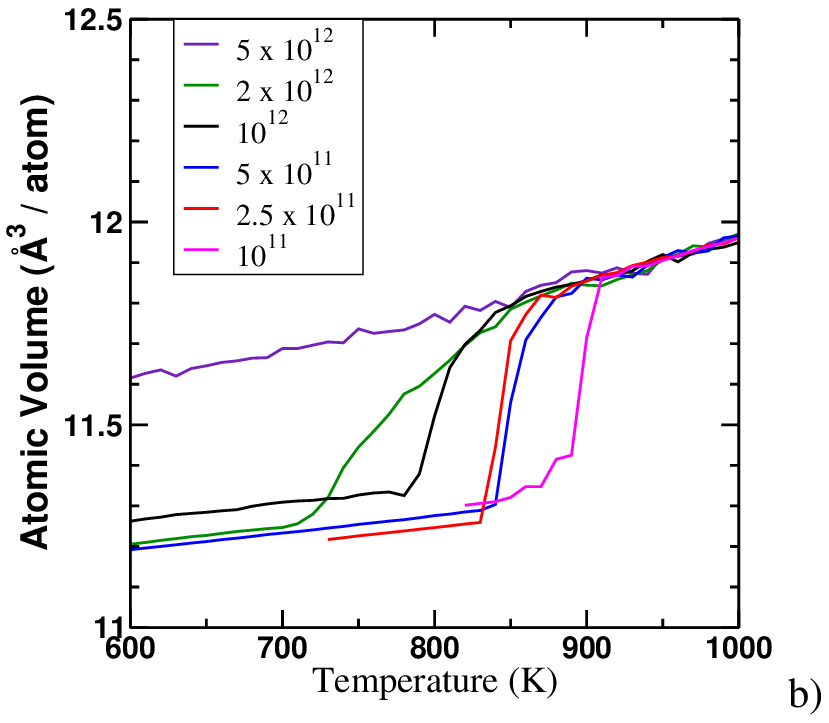}
\caption
{Figure a) and b) represent the atomic energy and atomic volume as a 
function of temperature for cooling rates between 10$^{11}$ and 5 x
10$^{12}$ K/s, respectively.}
\label{fig:volergsml}
\end{figure}

In Fig. \ref{fig:volergsml}, we present the average atomic energies
and atomic volumes as a function of temperature for various quench
rates. For quench rates below 5 x 10$^{12}$ K/s a rapid decrease in
energy and volume is seen in the region of temperature between 700-900
K. The equilibrium freezing temperature is 1540K\cite{che02}. 
The decrease occurs over a smaller range of temperature as the
quench rate is lowered.  In contrast, for quench rates of 5 x
10$^{12}$ K/s and higher (not shown) only a subtle change of slope in
the energy or volume is seen.  We attribute the rapid decrease in
energy and volume to crystallization and the change of slope to
solidification into an amorphous state.  Below the transition, the
curves for both energy and volume are parallel, but offset from each
other.  Thus the specific heat and thermal expansion of the
crystallized material are independent of quench rate (which determines
local structure) but the actual energy and volume differ primarily due
to quenched defects (vacancies) in the solid.  Above the transition
region note that the curves all lie on top of each other.  The
temperature derivative of the liquid energy (specific heat) and volume
(thermal expansion) are independent of quench rate.  Hence the local
structure in the liquid is not affected by the quench rate.

\begin{figure}
\includegraphics{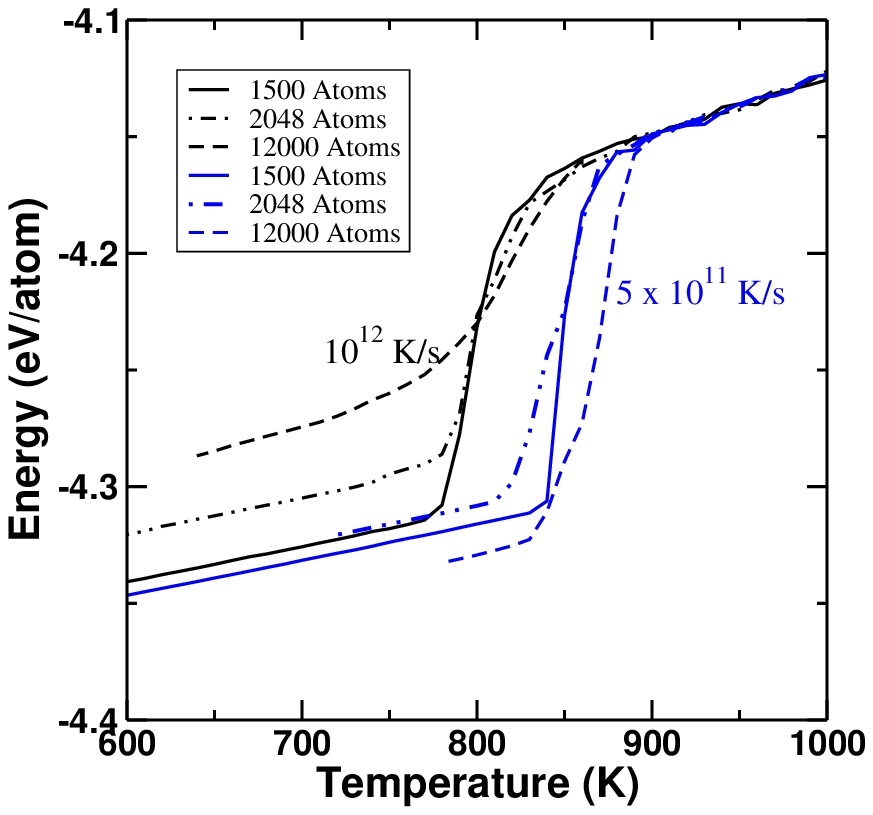}
\caption
{Comparison between a large system (12000 atoms), a medium system (2048
  atoms) and a small system 
(1500 atoms) of  the atomic energy as a function of temperature.  
The cooling rates shown are 5 x 10$^{11}$ and 10$^{12}$ K/s. The 
curves are for a single run.}

\label{fig:volerglgsml}
\end{figure}

To determine whether there are any important finite size effects
we compare the results in systems containing 12000, 2048
and 1500 atoms.  The resulting energy curves are
shown in Fig. \ref{fig:volerglgsml}.  The point where the slope begins
to increase as the temperature decreases indicates the onset of
crystallization. Nucleation occurs at a slightly
higher temperature for a given quench rate as the system size increases. 
This implies that 
the system size dependence for nucleation times is as
expected\cite{gunt} and finite size effects are not evident.

Analysis of the resulting microstructure showed that for
some conditions the small systems are polycrystalline and others were
single crystals. Similarly the large systems exhibit both
polycrystalline and single crystalline states. The distribution of  
the polycrystalline domains is similar in all sizes examined. In
addition the CNA as well as visual inspection(to be discussed
below) show similar patterns of nucleation. We thus conclude
that system size plays no important role in the resulting microstructure.

\begin{figure}
\includegraphics{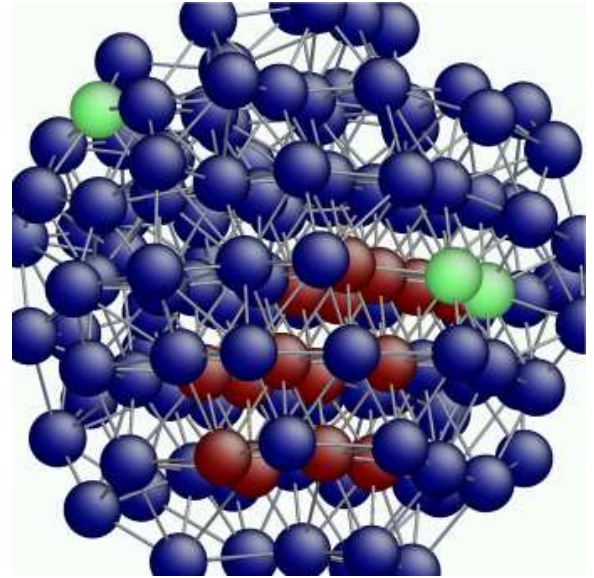}
\caption
{A 3D representation of the object found at $t_{o}$. The droplet
  consists of an fcc/hcp core surrounded by a halo consisting of
  planes with an in plane triangular structure. There are $80$ atoms in
  the droplet. The red(green) ball represents an fcc(hcp) atom while
  the blue balls are in a planar structure.}

\label{fig:3}
\end{figure}

\begin{figure}
\includegraphics{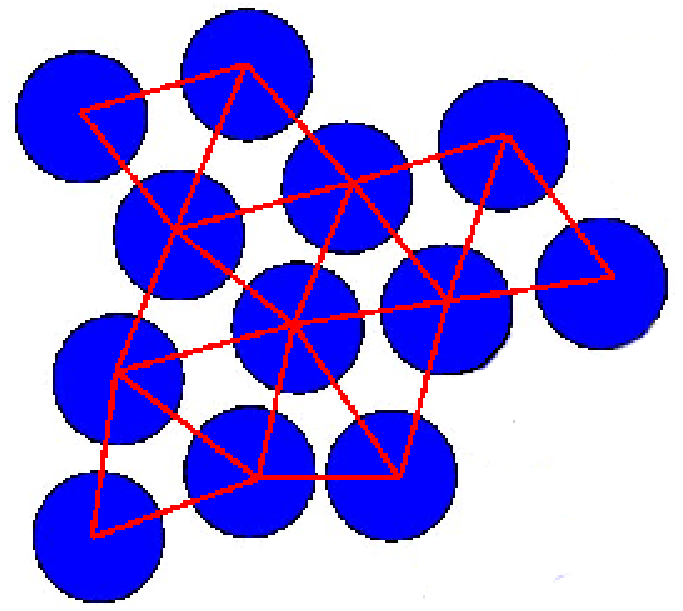}
\caption
{A 2D representation of one plane in the saddle point droplet. The droplet
  consists of 4 randomly stacked planes with an in plane triangular
  order. There are $60$ atoms in the droplet}
\label{fig:4}
\end{figure}

Using CNA and visual inspection we analyze the nucleation process
in supercooled liquid Ni.
Upon cooling for both
the large system (12,000 atoms) and the small system (1500 atoms),
a cluster of atoms having the fcc structure and the hcp structure {\it
  appear} to nucleate and grow at approximately $t_{o}=1.36$ nanoseconds
into the quench. We use {\it appear} since we are looking for the
saddle point object calculated in the theory.\cite{kle01}
 After some time into the growth of this cluster
the fcc particles dominate the structure. One  implication of this
result is that the argument of Alexander and McTague\cite{ale78},
does not apply to this particular system\cite{kle01}. This argument states
that in systems where the densities of the liquid and solid are similar
the solid phase has a  bcc structure. In addition the
prediction of icosahedral structures in supercooled liquids as
suggested by Frank\cite{fra52} (Lennard-Jones), J\'onsson {\it et
  al.}\cite{jon88} (Lennard-Jones, LJ), and Ramprasad {\it et
  al.}\cite{ram93} (Embedded Atom Method, EAM) is not observed.  We
attribute these differences to the spherical nature of the LJ and EAM
potentials, which do not include angular dependence included in the 
MEAM potential.  This characteristic appears to prevent the formation
of icosahedral ordering in pure nickel.

As discussed above, the KL theory \cite{kle86,kle01} predicts
stacked triangular planes or bcc for the droplet structure. The KL theory
also predicts the development
of a classical fcc/hcp core during
the growth phase. Therefore we look visually for a halo surrounding the
fcc/hcp structure described above. The possibility of a bcc halo was
excluded by the CNA analysis so we search for stacked planes with a
triangular structure. These were found as illustrated in Fig. \ref{fig:3}.
We then visually examine the system at 18 ps prior to the configuration at
$t_{o}$ to investigate the possibility that there is a droplet
consisting of only stacked triangular planes with no fcc/hcp
core. This configuration is also found and is illustrated in
Fig. \ref{fig:4}. These droplets contain $\sim 60$ atoms. Finally, we
investigate which of
these structures; stacked triangular planes with, or without, the
core correspond to the saddle point droplet predicted by the
theory.\cite{kle86,kle01} This is done via an intervention technique.
We take the system at a time which we consider it possible
that a nucleation droplet exists in the sense of a saddle point
solution. We then make $N$ copies of the system, scramble the
velocities keeping the temperature fixed, and then restart each
copy. In the $N\rightarrow \infty$ limit a saddle point implies that
half of the copies will continue to grow and half will disappear.\cite{mmr,mon}
 If the droplet is in the growth phase then
$N^{\prime}>>N/2$ droplets will grow. If the copies are made prior to
the saddle point droplet then $N^{\prime\prime}>>N/2$ droplets will
disappear. Using the configuration at $t_{o}=1.36$ nanoseconds we find that
in all of the copies the droplet grows. We also find that the
same result is obtained for configurations chosen at 1.5 and 11.5 ps
after $t_{o}$. Copies made at 28.5 ps prior to $t_{o}$ all contain
droplets that disappear. However, for configurations chosen at 8.5 and
18.5 ps prior to $t_{o}$ half of the droplets in the copies disappear
and half grow. Since we used $N=10$ some spread in the location of the
saddle point droplet was expected. In addition, the saddle point
flattens at the top near a spinodal.\cite{uk85} Even though there was
some spread in the location of the saddle point droplet it is important
to note that the droplets at these two times consist only of
randomly stacked planes with an in plane triangular structure and do
not contain an fcc/hcp core.

In conclusion, our study shows
that the local structure of the initial (saddle point) nucleation droplet 
consists of randomly stacked planes with a triangular ordering. During the
initial growth phase a dense core develops which has both fcc and hcp atoms.
Further into the growth phase the core becomes dominated by fcc atoms.
This nucleation sequence is in direct contradiction of the argument of 
Alexander and McTague\cite{ale78,kle01}
and appears to exhibit the non-classical behavior that Klein and
Leyvraz\cite{kle86, kle01} predicted. This result has important
implications for the calculation of nucleation rates and the
understanding of the morphology in either the stable or metastable
state that has been formed via nucleation. It also gives additional
evidence that the spinodal, or limit of metastability of super-cooled
liquids may play an important role in determining the local structure of
a critical droplet. Finally, non-classical droplets have been
found in Lennard-Jones fluids\cite{yang88, yang90} and colloids,\cite{gasser01} 
however this is the first evidence for non-classical nucleation
influenced by a spinodal in a model of a molecular material such as a metal.

\bibliographystyle{apsrev}

\end{document}